\documentclass{ws-procs9x6}

\begin{document}

\title{New physics at low energies and 
\\
dark matter-dark energy transmutation}

\author{E. I. Guendelman and A.  B.  Kaganovich }

\address{Physics Department, Ben Gurion University of the Negev, 
\\
Beer Sheva
84105, Israel
\\
E-mail: guendel@bgumail.bgu.ac.il ; alexk@bgumail.bgu.ac.il}

\maketitle

\abstracts{
  A field theory is proposed where the regular fermionic matter and the dark
fermionic matter can be different states of the same "primordial" fermion fields.
In regime of the fermion densities typical for normal particle physics, the
primordial fermions split into three families identified with regular fermions.
When fermion energy density becomes comparable with dark energy density, the theory
allows transition to new type of states. The possibility of such Cosmo-Low Energy Physics
(CLEP) states is demonstrated by means of solutions of the field theory
equations describing FRW universe filled with homogeneous scalar field and 
uniformly distributed nonrelativistic neutrinos. Neutrinos in CLEP state
are drawn into
cosmological expansion by means of dynamically changing their own parameters.
One of the features of the fermions in CLEP state is that in the late time
universe their masses
increase as $a^{3/2}$ ($a=a(t)$ is the scale factor). The energy
density of the cold dark matter consisting of neutrinos in CLEP state scales as
a sort of dark energy; this cold dark matter possesses negative pressure and for the
late time universe its equation of state approaches that of the cosmological
constant. The total energy density of such universe is less than it would be in
the universe free of fermionic matter at all. 
}

\section{Main ideas of the Two Measures Theory and the scale
invariant model}

The Two Measures Theory (TMT) is a generally coordinate invariant theory
where the action has the form
\begin{equation}
    S = \int L_{1}\Phi d^{4}x +\int L_{2}\sqrt{-g}d^{4}x
\label{S}
\end{equation}
 including two Lagrangians $ L_{1}$ and $L_{2}$ and two 
measures of the volume elements ($\Phi d^{4}x$ and $\sqrt{-g}d^{4}x$
respectively). One is the usual measure of integration
$\sqrt{-g}$ in the 4-dimensional space-time manifold equipped by the metric
 $g_{\mu\nu}$. Another is also a scalar density built of four scalar
fields
$\varphi_{a}$ ($a=1,2,3,4$), \,
$\Phi
=\varepsilon^{\mu\nu\alpha\beta}\varepsilon_{abcd}\partial_{\mu}\varphi_{a}
\partial_{\nu}\varphi_{b}\partial_{\alpha}\varphi_{c}
\partial_{\beta}\varphi_{d}$.
It is assumed that the Lagrangians $ L_{1}$ and $L_{2}$  are functions
of the matter
fields, the dilaton field, the metric, the connection  (or spin-connection )
 but not of the
"measure fields" $\varphi_{a}$. 
Varying $\varphi_{a}$, we get
$B^{\mu}_{a}\partial_{\mu}L_{1}=0 $ where  
$B^{\mu}_{a}=\varepsilon^{\mu\nu\alpha\beta}\varepsilon_{abcd}
\partial_{\nu}\varphi_{b}\partial_{\alpha}\varphi_{c}
\partial_{\beta}\varphi_{d}$.
Since
$Det (B^{\mu}_{a})
= \frac{4^{-4}}{4!}\Phi^{3}$ it follows that if $\Phi\neq 0$,
$ L_{1}=sM^{4} =const$
where $s=\pm 1$ and $M$ is a constant of integration with the dimension of
mass.
Important feature of TMT that is responsible for many
interesting and desirable results of the field theory models studied 
so far\cite{GK1}-\cite{GK6} (see also \cite{proceed-2001})
 consists of the assumption that all fields, including 
also metric, connection (or vierbein and spin-connection) and the
 measure fields $\varphi_{a}$ are independent dynamical variables. 

As it has been shown earlier\cite{}, in TMT there is no a need to postulate
the existence of three species for each type of fermions (like three neutrinos,
three charged leptons, etc.) but rather this is achived as a dynamical effect
of TMT in normal particle physics conditions.  
The matter content of our model includes the dilaton scalar 
field $\phi$, two 
so-called primordial fermion fields (the neutrino primordial field $\nu$
 and the electron primordial field $E$) and electromagnetic field $A_{\mu}$.
Generalization to the non-Abelian 
gauge models including Higgs fields and quarks is straightforward\cite{GK5}.
To simplify the presentation of the ideas we ignore also the chiral 
properties of neutrino; this
can be done straightforward and does not affect the main results. 

Keeping the general structure 
(\ref{S}), it is convenient to represent the action in the following 
form:
\begin{eqnarray}
S &=& \int d^{4}x e^{\alpha\phi /M_{p}}
(\Phi +b\sqrt{-g})\left[-\frac{1}{\kappa}R(\omega ,e)
+\frac{1}{2}g^{\mu\nu}\phi_{,\mu}\phi_{,\nu}\right]
\nonumber\\
&-&\int d^{4}x e^{2\alpha\phi /M_{p}}[\Phi V_{1} +\sqrt{-g}
V_{2}]-\int d^{4}x\sqrt{-g}
\frac{1}{4}g^{\alpha\beta}g^{\mu\nu}F_{\alpha\mu}F_{\beta\nu}
\nonumber\\
&+&\int d^{4}x e^{\alpha\phi /M_{p}}(\Phi +k\sqrt{-g})\frac{i}{2}
\sum_{i}\overline{\Psi}_{i}
\left(\gamma^{a}e_{a}^{\mu}\overrightarrow{\nabla}_{\mu}^{(i)}
-\overleftarrow{\nabla}_{\mu}^{(i)}
\gamma^{a}e_{a}^{\mu}\right)\Psi_{i}
\nonumber\\
&-&\int d^{4}x e^{\frac{3}{2}\alpha\phi /M_{p}}
\left[(\Phi +h_{\nu}\sqrt{-g})\mu_{\nu}\overline{\nu}\nu
+(\Phi +h_{E}\sqrt{-g})\mu_{E}\overline{E}E
\right]
\label{totaction}
\end{eqnarray}
where where $\Psi_{i}$ ($i=\nu , E$) is the
general notation for the primordial fermion fields $\nu$ and $E$,
 $V_{1}$ and $V_{2}$ are constants, 
$F_{\alpha\beta}=\partial_{\alpha}A_{\beta}-
\partial_{\beta}A_{\alpha}$, \quad
   $\mu_{\nu}$ and
$\mu_{E}$ are  the mass parameters,
$\overrightarrow{\nabla}_{\mu}^{(\nu)}=\overrightarrow{\partial}_{\mu}+
\frac{1}{2}\omega_{\mu}^{cd}\sigma_{cd}$,
$\overrightarrow{\nabla}^{(E)}_{\mu}=\overrightarrow{\partial}_{\mu}+
\frac{1}{2}\omega_{\mu}^{cd}\sigma_{cd}+ieA_{\mu}$;
$R(\omega ,e) =e^{a\mu}e^{b\nu}R_{\mu\nu ab}(\omega)$ is
the scalar curvature,
 $e_{a}^{\mu}$ and
$\omega_{\mu}^{ab}$ are the vierbein  and spin-connection;
$g^{\mu\nu}=e^{\mu}_{a}e^{\nu}_{b}\eta^{ab}$ and
$R_{\mu\nu ab}(\omega)=\partial_{\mu}\omega_{\nu ab}   
+\omega_{\mu a}^{c}\omega_{\nu cb}
-(\mu\leftrightarrow\nu)$; constants $b, \, k, \, h_{i}$
are dimensionless parameters.

The action (\ref{totaction}) is invariant
under
the global scale transformations
\begin{eqnarray}
    e_{\mu}^{a}\rightarrow e^{\theta /2}e_{\mu}^{a}, \quad
\omega^{\mu}_{ab}\rightarrow \omega^{\mu}_{ab}, \quad
\varphi_{a}\rightarrow \lambda_{a}\varphi_{a}\quad
where \quad \Pi\lambda_{a}=e^{2\theta}
\nonumber
\\
A_{\alpha}\rightarrow A_{\alpha}, \quad
\phi\rightarrow \phi-\frac{M_{p}}{\alpha}\theta ,\quad
\Psi_{i}\rightarrow e^{-\theta /4}\Psi_{i}, \quad
\overline{\Psi}_{i}\rightarrow  e^{-\theta /4} \overline{\Psi}_{i}.
\label{stferm}
\end{eqnarray}

One can show that except for a few special choices providing positivity of the
energy and the right chiral structure in the Einstein frame,
Eq.(\ref{totaction}) describes {\it
the most
general TMT action satisfying the formulated above symmetries}.

\section{Constraint and equations of motion in the
Einstein frame}

Variation of
the measure fields $\varphi_{a}$ with the condition
$\Phi\neq 0$ leads, as we have already seen
in Sec.2, to the equation $ L_{1}=sM^{4}$  
where $L_{1}$ is now defined, according to  Eq. (\ref{S}), as
the part of the integrand of the action (\ref{totaction})
coupled to the measure $\Phi$. 
The appearance of a nonzero integration
constant $sM^{4}$ spontaneously breaks the scale invariance 
(\ref{stferm}). 
One can see that the measure $\Phi$ degrees of freedom 
appear in all the equations of motion only through dependence
on the scalar 
field $\zeta \equiv\Phi/\sqrt{-g}$.
In particular, the gravitational and all matter
fields equations of motion include noncanonical terms 
proportional to $\partial_{\mu}\zeta$.
It turns out that with the set of the new variables ($\phi$ and 
$A_{\mu}$ remain the same)
\begin{equation}
\tilde{e}_{a\mu}=e^{\frac{1}{2}\alpha\phi/M_{p}}(\zeta
+b)^{1/2}e_{a\mu}, \quad
\Psi^{\prime}_{i}=e^{-\frac{1}{4}\alpha\phi/M_{p}}
\frac{(\zeta +k)^{1/2}}{(\zeta +b)^{3/4}}\Psi_{i}
\label{ctferm}
\end{equation}
which we call the Einstein frame,
 the spin-connections become those of the 
Einstein-Cartan space-time and the noncanonical terms proportional to 
$\partial_{\mu}\zeta$ disappear from all equations of motion. Since  
$\tilde{e}_{a\mu}$,
$\nu^{\prime}$ and $E^{\prime}$ are invariant under the scale 
transformations
(\ref{stferm}), spontaneous breaking of the scale symmetry (\ref{stferm})
(by means of equation $L_{1}=sM^{4}$) is reduced in the new variables to the
{\it spontaneous breaking of the shift symmetry} 
$\phi\rightarrow\phi +const$.

The gravitational 
equations 
in the Einstein frame take the standard GR form
$G_{\mu\nu}(\tilde{g}_{\alpha\beta})=\frac{\kappa}{2}T_{\mu\nu}^{eff}$
where
\begin{equation}
T_{\mu\nu}^{eff}=K_{\mu\nu}
+\tilde{g}_{\mu\nu}V_{eff}(\phi ;\zeta)+T_{\mu\nu}^{(em)}
+T_{\mu\nu}^{(f,can)}+T_{\mu\nu}^{(f,noncan)}
 \label{Tmn}
\end{equation}
Here $K_{\mu\nu}=\phi_{,\mu}\phi_{,\nu}-\frac{1}{2}
\tilde{g}_{\mu\nu}\tilde{g}^{\alpha\beta}\phi_{,\alpha}\phi_{,\beta}$,
\quad
$G_{\mu\nu}(\tilde{g}_{\alpha\beta})$ is the
Einstein tensor
in the Riemannian space-time with the metric
$\tilde{g}_{\mu\nu}$; the function $V_{eff}(\phi ;\zeta)$ has the form 
\begin{equation}
V_{eff}(\phi ;\zeta)=
\frac{b\left(sM^{4}e^{-2\alpha\phi/M_{p}}+V_{1}\right)-V_{2}}{(\zeta
+b)^{2}};
\label{Veff1}
\end{equation}
$T_{\mu\nu}^{(em)}$ is the canonical energy momentum tensor for the
electromagnetic field;
$T_{\mu\nu}^{(f,can)}$ is the canonical energy momentum tensor for
(primordial) fermions $\nu^{\prime}$ and $E^{\prime}$ in
curved space-time
including also standard electromagnetic interaction of $E^{\prime}$.
$T_{\mu\nu}^{(f,noncan)}$ is the {\em noncanonical} contribution
of the fermions into the energy momentum tensor 
\begin{equation}
 T_{\mu\nu}^{(f,noncan)}=-\tilde{g}_{\mu\nu}\sum_{i}F_{i}(\zeta)
\overline{\Psi^{\prime}}_{i}\Psi^{\prime}_{i}
\equiv \tilde{g}_{\mu\nu}\Lambda_{dyn}^{(ferm)}
\label{Tmn-noncan}
\end{equation}
where $i=\nu^{\prime},E^{\prime}$ and
\begin{equation}
F_{i}(\zeta)\equiv
\frac{\mu_{i}}{2(\zeta +k)^{2}(\zeta +b)^{1/2}}
[\zeta^{2}+(3h_{i}-k)\zeta +2b(h_{i}-k)+kh_{i}].
 \label{Fizeta1}
\end{equation}
The structure of 
$T_{\mu\nu}^{(f,noncan)}$ shows that it is
originated by fermions but  behaves as
 a sort of  variable cosmological constant.
This is
why we will refer to it as {\it dynamical fermionic $\Lambda$  
 term}
$\Lambda_{dyn}^{(ferm)}$. One has to 
emphasize
the substantial difference of the way $\Lambda_{dyn}^{(ferm)}$
 emerges here as compared to the models of the
condensate cosmology (see for example Ref.\cite{cond}).
As we will see in the next sections, 
$\Lambda_{dyn}^{ferm}$ becomes negligible in gravitational experiments
with observable matter.
However it may be  very important for some
astrophysics and cosmology problems.

The dilaton $\phi$ field equation in the new variables
reads
\begin{equation}
\Box\phi -\frac{\alpha}{M_{p}(\zeta +b)} 
\left[sM^{4}e^{-2\alpha\phi/M_{p}}-\frac{(\zeta -b)V_{1}+2V_{2}}{\zeta
+b}\right]=
-\frac{\alpha }{M_{p}}\sum_{i}F_{i}\overline{\Psi^{\prime}_{i}}\Psi^{\prime}_{i}
\label{phief+ferm1}
\end{equation}
where $\Box\phi =(-\tilde{g})^{-1/2}\partial_{\mu}
(\sqrt{-\tilde{g}}\tilde{g}^{\mu\nu}\partial_{\nu}\phi)$.

Equations for the primordial leptons 
in terms of the
variables (\ref{ctferm}) take the standard form of fermionic equations
in the Einstein-Cartan space-time  where the standard electromagnetic
interaction 
 presents also. All the novelty consists of the form of
the $\zeta$ depending "masses" $m_{i}(\zeta)$ of the primordial fermions
$\nu^{\prime}$, $E^{\prime}$:
\begin{equation}
m_{i}(\zeta)=
\frac{\mu_{i}(\zeta +h_{i})}{(\zeta +k)(\zeta +b)^{1/2}}
\qquad i=\nu^{\prime},E^{\prime}.
 \label{muferm1} 
\end{equation} 
It should be noticed that change of variables we have performed
by means of Eq.(\ref{ctferm}) provide also a conventional form
of the covariant conservation law of fermionic current
$j^{\mu}=\overline{\Psi}^{\prime}\gamma^{a}\tilde{e}_{a}^{\mu}\Psi^{\prime}$. 

The scalar field $\zeta$ in the above equations is defined
by the constraint which is the consistency condition of
equations of motion, see for example\cite{proceed-2001}.
In the Einstein frame
(\ref{ctferm}) the constraint takes the form
\begin{equation}
-\frac{1}{(\zeta +b)^{2}}\left\{(\zeta
-b)\left[sM^{4}e^{-2\alpha\phi/M_{p}}+
V_{1}\right]+2V_{2}\right\}=
\sum_{i}F_{i}\overline{\Psi^{\prime}_{i}}\Psi^{\prime}_{i}.
\label{constraint3}
\end{equation}

Generically the constraint (\ref{constraint3}) determines $\zeta$ as 
a very complicated 
function of $\phi$, \,
$\overline{\nu^{\prime}}\nu^{\prime}$ and $\overline{E^{\prime}}E^{\prime}$ . 
However, there are  a few very 
important particular situations where the theory allows exact
solutions of great interest\cite{GK4},\cite{GK5}.

\section{Some limiting cases of physical interest}

In a typical particle physics
situation, say detection of a single fermion, 
the measurement implies a 
localization of the fermion which is expressed in developing a very large 
value of  $|\overline{\Psi^{\prime}}\Psi^{\prime}|$.
According to the constraint (\ref{constraint3}) this is
possible if
$F_{i}(\zeta)\approx 0, \, i=\nu^{\prime},E^{\prime}$
(which gives two constant solutions for $\zeta$)
or
$\zeta \approx -b$.
These solutions allow to describe the effect of splitting 
of the primordial fermions into three generations of the
regular fermions (for details see \cite{GK4},\cite{GK5},cite{proceed-2001}).
 It is interesting also
that for the first two generations (which we associate with the solutions
where $F_{i}(\zeta)\approx 0$) their coupling to the dilaton $\phi$
is automatically strongly suppressed, as it follows from Eq.(\ref{phief+ferm1}),
which provides a solution of the fifth force problem (about the role of the symmetry
$\phi\rightarrow \phi +const$ see Ref.\cite{Carroll})  

In the case of the complete absence of massive fermions the constraint determines
$\zeta$ as the function of $\phi$:
$\zeta = b-2V_{2}/
(V_{1}+sM^{4}e^{-2\alpha\phi/M_{p}})$.
The effective potential of the scalar field $\phi$ results then from Eq.(\ref{Veff1})
\begin{equation} 
V_{eff}^{(0)}(\phi)\equiv V_{eff}(\phi;\zeta)|_{\overline{\psi^{\prime}}\psi^{\prime}=0}
=\frac{(V_{1}+sM^{4}e^{-2\alpha\phi/M_{p}})^{2}}
{4[b(V_{1}+sM^{4}e^{-2\alpha\phi/M_{p}})-V_{2}]}.
\label{Veffvac} 
\end{equation}

Assuming $bV_{1}>V_{2}$ and $s=+1$ we see that the asymptotic
(as $\phi\rightarrow\infty$) value of $V_{eff}^{(0)}$
is the positive
cosmological constant
$\Lambda^{(0)}
=\frac{V_{1}^{2}}   
{4(bV_{1}-V_{2})}$. If $2V_{2}>bV_{1}>V_{2}$
then $V_{eff}^{(0)}$ has the absolute minimum 
$V_{eff,min}^{(0)}=V_{2}/b^{2}$
at $\phi =\phi_{min}=(M_{p}/2\alpha)\ln[bM^{4}/(2V_{2}-bV_{1})]$.

\section{Cosmo-Low Energy Physics states }
It turns out that besides the normal fermion vacuum
where the fermion contribution to the constraint is totally
negligible, 
TMT predicts possibility of so far unknown
 states which can be realized,  for example,
in astrophysics and cosmology. Let us study a toy model\cite{GK6}
 where 
in addition to the 
homogeneous scalar field $\phi$,  the spatially flat
universe is filled also with uniformly 
distributed nonrelativistic neutrinos as a model of  dark matter.
 Spreading of the neutrino
wave packets during their free motion lasting a long time yields
extremely small values of 
 $\overline{\Psi}^{\prime}\Psi^{\prime}= u^{\dagger}u$
($u$ is the large component of the Dirac spinor $\Psi^{\prime}$).
There is a solution where the decaying fermion 
contribution $u^{\dagger}u\sim \frac{const}{a^{3}}$ to the 
constraint is compensated by
approaching $\zeta\rightarrow -k$. Then solving (\ref{constraint3})
for $\zeta$ we have to take into account both sides of the 
constraint.
 After 
averaging over typical cosmological scales (resulting in 
the Hubble low), the constraint (\ref{constraint3})  reads
\begin{equation}
-(k+b)\left(sM^{4}e^{-2\alpha\phi/M_{p}}+V_{1}\right)
+2V_{2}+ 
(b-k)^{2}\frac{n^{(\nu)}_{0}}{a^{3}}
F_{\nu}(\zeta)|_{\zeta\approx -k}
=0
\label{constr-k}
\end{equation} 
where  $F_{\nu}(\zeta)|_{\zeta\approx -k}=
\mu_{\nu}(h_{\nu}-k)(b-k)^{1/2}(\zeta +k)^{-2}+O((\zeta +k)^{-1})$
 and $n^{(\nu)}_{0}$ is a constant
determined by the total number of the cold neutrinos.

Cosmological equations 
are then as  following
\begin{equation}
\left(\frac{\dot{a}}{a}\right)^{2}=\frac{1}{3M_{p}^{2}}\left[\rho_{\phi}
+\rho_{clep}\right]
\label{FRW-eq1}
\end{equation}
\begin{equation}
\ddot{\phi}+3\frac{\dot{a}}{a}\dot{\phi}
+\frac{2\alpha k}{(b-k)^{2}M_{p}}M^{4}e^{-2\alpha\phi/M_{p}}
+O\left((\zeta +k)e^{-2\alpha\phi/M_{p}}\right)
=0
\label{d.e.-eq+constr}
\end{equation}
where 
\begin{equation}
\rho_{\phi}=\frac{1}{2}\dot{\phi}^{2}+
\frac{bV_{1}-V_{2}}{(b-k)^{2}}
+\frac{b}{(b-k)^{2}}sM^{4}e^{-2\alpha\phi/M_{p}}
+O(\zeta +k),
\label{rho-d.e.}
\end{equation}
\begin{equation}
\rho_{clep}=\left[\frac{\mu_{\nu}(h_{\nu}-k)}
{(\zeta +k)(b-k)^{1/2}}-
F_{\nu}(\zeta)|_{\zeta\approx -k}\right]
\frac{n^{(\nu)}_{0}}{a^{3}}
\label{rho-d.m.}
\end{equation}
The corrections $O(\zeta +k)$ behave as
$O(\zeta +k)\propto a^{-3/2}$ and it turns out that
they are negligible if no fine
tuning of the parameters of the theory will be done. 
We choose $s=+1$ and assume here that 
$V_{1}>0, \enspace V_{2}>0 \quad and \quad
b>0, \enspace k<0, \enspace h_{\nu}<0, \enspace h_{\nu}-k<0,
 \enspace b+k<0 $.
 In the case we work
with a positive fermion energy solution, in order to provide
positivity of the effective neutrino mass (see Eq.(\ref{muferm1}))
 in the CLEP
state, we should consider the regime where 
$\zeta =-k-\varepsilon$, $\varepsilon >0$. Then 
$F_{\nu}(\zeta)|_{\zeta\approx -k}<0$
and both of the terms in Eq.(\ref{rho-d.m.}) are positive.
The first term in Eq.(\ref{rho-d.m.}) results from the 
canonical part 
$T_{00}^{(\nu,can)}$ of the neutrino 
energy-momentum
tensor after 
making use of the equations for
 neutrino field,
neglecting the terms proportional
to 3-momenta of the neutrinos and 
averaging. The second term in Eq.(\ref{rho-d.m.}) comes
from the dynamical fermionic
$\Lambda_{dyn}^{(ferm)}$ term.

Neglecting in (\ref{rho-d.m.}) terms of the order
of $(\zeta +k)^{-1}$ as compared
to the terms of the order of $(\zeta +k)^{-2}$ 
and  using again the constraint (\ref{constr-k})
we obtain for the  pressure and density of the uniformly 
distributed neutrino in the CLEP state 
 \begin{equation}
P_{clep}=-\rho_{clep}=\frac{2V_{2}+|b+k|V_{1}}{(b-k)^{2}}
+\frac{|b+k|}{(b-k)^{2}}
M^{4}e^{-2\alpha\phi/M_{p}}
\label{rho-dm-de}
\end{equation}
which is typical for the dark energy sector including both 
a cosmological constant and an exponential $\phi$-potential
(compare (\ref{rho-dm-de}) with (\ref{rho-d.e.})). 
The accuracy of this approximation grows as $a(t)\rightarrow\infty$.

The total energy density and
the total pressure  
(including both the scalar field $\phi$ and neutrinos in 
CLEP state ) in the framework
 of the explained
 above approximations can be represented in the form
\begin{equation}
\rho^{(total)}_{dark}\equiv\rho_{\phi}+\rho_{clep}
 =\frac{1}{2}\dot{\phi}^{2}+U_{dark}^{(total)}(\phi)
\label{tot-rho-nu}
\end{equation}   
\begin{equation}
P^{(total)}_{dark}\equiv P_{\phi}+P_{clep}
=\frac{1}{2}\dot{\phi}^{2}-U_{dark}^{(total)}(\phi),
\label{tot-p-nu}
\end{equation}
where the effective potential 
$U_{dark}^{(total)}(\phi)=\Lambda +V_{quint}(\phi)$,
where 
\begin{equation}
\Lambda =\frac{V_{2}+|k|V_{1}}{(b-k)^{2}}, \qquad
V_{quint}(\phi)=\frac{|k|}{(b-k)^{2}}M^{4}e^{-2\alpha\phi/M_{p}}.
\label{Ueff-phi}
\end{equation}
This means that the evolution 
of the late time universe in the state with
$\zeta\approx -k$
proceeds as it would be in the standard field theory
 model (non-TMT) including
{\it both the cosmological constant
 and the quintessence-like field $\phi$ with the exponential
potential}. Note that to provide the observable 
energy densities (for example, 
$\Lambda\sim\rho_{crit}$, where $\rho_{crit}$ is the present
day critical energy density) there is no need of fine tuning
of the dimensionfull parameters  $V_{1}$ and $V_{2}$ but    
instead one can assume that the dimensionless parameters
$b$, $k$ are very large.

It is very interesting to compare the effective potential
$V_{eff}^{(0)}(\phi)$, Eq.(\ref{Veffvac}),
predicted for the universe filled only with the homogeneous scalar field
 (for short, a state "absent of fermions"), on the one hand,
 with the effective dark sector potential $U_{dark}^{(total)}(\phi)$
 for the universe filled both with the homogeneous scalar field 
and with the uniformly distributed nonrelativistic neutrinos
 (for short, "CLEP state"), on the other hand.
 The {\it remarkable result} consists
in the fact that if $bV_{1}>V_{2}$, which is needed for positivity
of $\Lambda^{(0)}$, then     
\begin{equation}
V_{eff}^{(0)}(\phi)-U_{dark}^{(total)}(\phi)\equiv
\frac{\left[\frac{b+k}{2}\left(V_{1}+M^{4}e^{-2\alpha\phi /M_{p}}\right)
-V_{2}\right]^{2}}
{4(b-k)^{2}\left[b\left(V_{1}+M^{4}e^{-2\alpha\phi /M_{p}}\right)
-V_{2}\right]}>0.
\label{L-L0}
\end{equation}
This means that (for the same value of $\dot{\phi}^{2}$)
{\it the universe in "the CLEP state" has a lower energy density
than the one in the  "absent of fermions" state}. One should 
emphasize that this result does not imply at all that $\rho_{clep}$
is negative.

For
illustration of what kind of solutions one can expect, let us take the
{\em particular value} for the parameter $\alpha$, namely
$\alpha =\sqrt{3/8}$.
Then in the framework of the explained above approximations,
 the cosmological equations allow the following analytic solution:  
 \begin{equation}
\phi(t)=\frac{M_{p}}{2\alpha}\varphi_{0}+
\frac{M_{p}}{2\alpha}\ln(M_{p}t),
\qquad
a(t)\propto t^{1/3}e^{\lambda t},
\label{a-sol-nu}
\end{equation}
where
\begin{equation}
\lambda =\frac{1}{M_{p}}\sqrt{\frac{\Lambda}{3}},
\qquad e^{-\varphi_{0}}=
\frac{2(b-k)^{2}M_{p}^{2}}{\sqrt{3}|k|M^{4}}\sqrt{\Lambda}.
\label{phi-0}
\end{equation}
The mass of the neutrino in such CLEP state increases
exponentially in time and its $\phi$ dependence is
double-exponential:
\begin{equation}
m_{\nu}|_{CLEP}\sim a^{3/2}(t)\sim
t^{1/2}e^{\frac{3}{2}\lambda t}\sim 
\exp\left[\frac{3\lambda e^{-\varphi_{0}}}{2M_{p}}
\exp\left(\frac{2\alpha}{M_{p}}\phi \right)\right].
\label{m-t-phi}
\end{equation}

\end{document}